\documentclass[11pt,twoside]{article}
\usepackage[pdftex]{graphicx}
\usepackage{amsmath}
\usepackage{amssymb}
\usepackage{cite}
\usepackage{braket}

\begin{document}
\newcommand{\pst}{\hspace*{1.5em}}

\newcommand{\lemark}{\em }

%\lhead[\fancyplain{\rigmark, {\em \lemark}}{\rigmark}]{\fancyplain{\rigmark, {\em \lemark}}{\lemark}}
%\chead{}\rhead[\fancyplain{}{\lemark}]{\fancyplain{}{\rigmark}}
%\plainfootrulewidth 0.4pt
\newcommand{\be}{\begin{equation}}
\newcommand{\ee}{\end{equation}}
\newcommand{\bm}{\boldmath}
\newcommand{\ds}{\displaystyle}
\newcommand{\bea}{\begin{eqnarray}}
\newcommand{\eea}{\end{eqnarray}}
\newcommand{\ba}{\begin{array}}
\newcommand{\ea}{\end{array}}
\newcommand{\arcsinh}{\mathop{\rm arcsinh}\nolimits}
\newcommand{\arctanh}{\mathop{\rm arctanh}\nolimits}
\newcommand{\bc}{\begin{center}}
\newcommand{\ec}{\end{center}}

\thispagestyle{plain}

\label{sh}

%\lfoot[\fancyplain{\ \\[1mm] \thepage}{\ \\[1mm]\thepage}]{\fancyplain{}{}}

\begin{center} {\Large \bf
\begin{tabular}{c}
H-theorem and Maxwell Demon\\ in Quantum Physics
\end{tabular}
 } \end{center}

\bigskip

\bigskip

\begin{center} {\bf
N.S.\ Kirsanov$^{1*}$, A.V.\ Lebedev$^{2,1}$, I.A.\ Sadovskyy$^3$, M.V.\ Suslov$^1$,\\
V.M.\ Vinokur$^3$, G.\ Blatter$^2$, and G.B.\ Lesovik$^1$
}\end{center}

\medskip

\begin{center}
{\it
$^1$Moscow Institute of Physics and Technology \\
141700, Institutskii
Per. 9, Dolgoprudny, Moscow Distr., Russian Federation

\smallskip

$^2$Theoretische Physik, Wolfgang-Pauli-Strasse 27, ETH Z\"{u}rich,
CH-8093 Z\"{u}rich, Switzerland
}

\smallskip

$^3$Materials Science Division, Argonne
National Laboratory, 9700 S. Cass Ave., Argonne, IL 60439, USA

\smallskip

$^*$nikita.kirsanov@phystech.edu\\
\end{center}

\begin{abstract}\noindent
The Second Law of Thermodynamics states that temporal evolution of an isolated system occurs with non-diminishing
entropy. In quantum realm, this holds for energy-isolated systems the evolution of which is described by the so-called unital
quantum channel. The entropy of a system evolving in a non-unital quantum channel can, in principle, decrease. We formulate
a general criterion of unitality for the evolution of a quantum system, enabling a simple and rigorous approach for finding and
identifying the processes accompanied by decreasing entropy in energy-isolated systems. We discuss two examples illustrating our
findings, the quantum Maxwell demon and heating-cooling process within a two-qubit system.
\end{abstract}

\medskip

\noindent{\bf Keywords:}
Quantum information, H-theorem, quantum mechanics, qubits, quantum Maxwell demon 

\section{Introduction}

\begin{figure}[h]
  \centerline{\includegraphics[width=310pt]{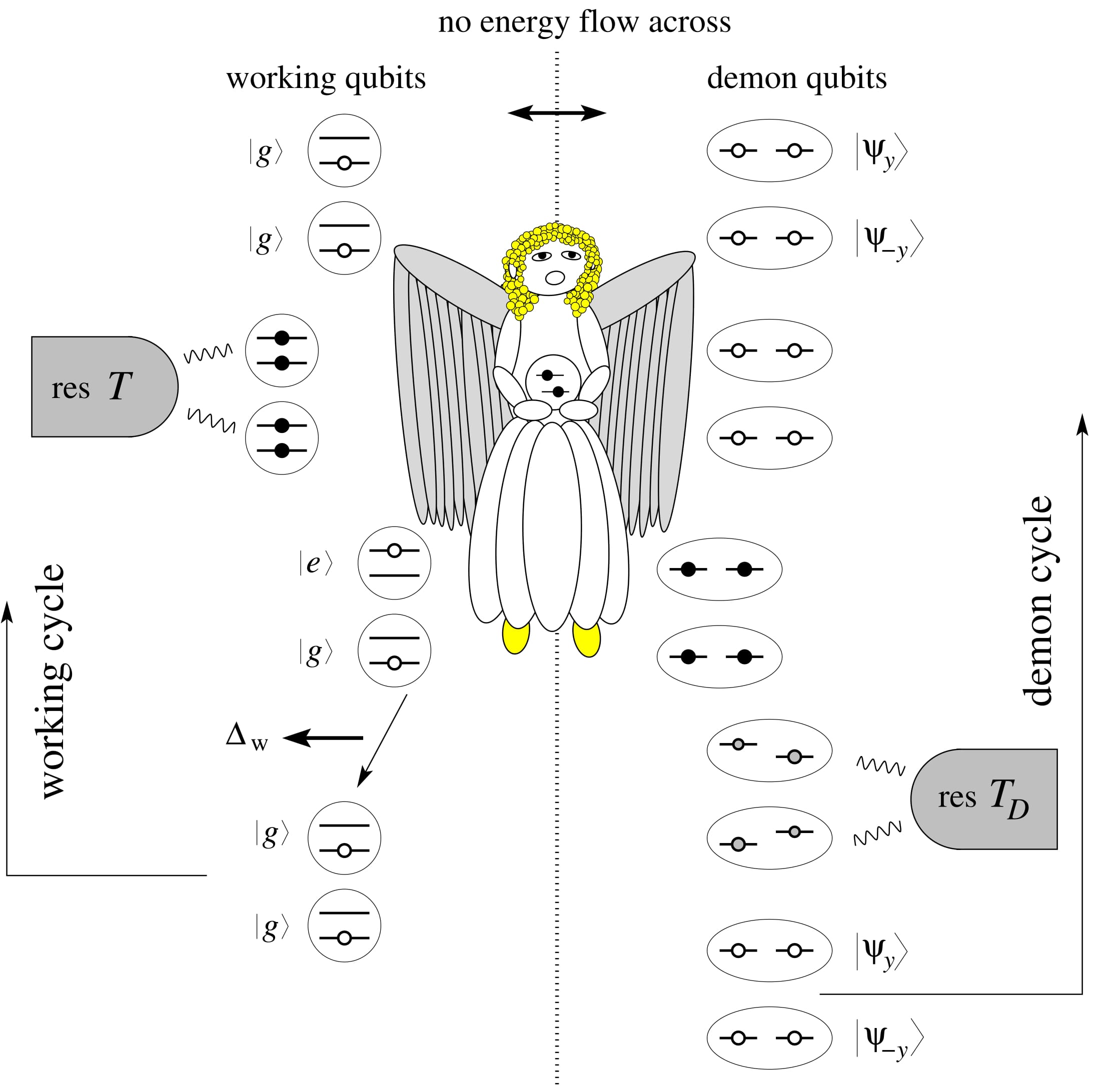}}
  \caption{Heat engine with separate energy- and entropy cycles 
  enabled by a quantum angel.}
  \label{fig1}
\end{figure}

Recently, building on the mathematical formalism of quantum information theory
(QIT), we have formulated a quantum H-theorem in terms of physical
observables\,\cite{Lesovik}. Consider a fixed-energy subspace $E$ of the
system's Hilbert space spanned by the orthonormal basis states
$\ket{\psi_{i,E}}$, $\hat{H}_S\ket{\psi_{i,E}}=E\ket{\psi_{i,E}}$, where the
index $i$ denotes all of the system's remaining non-energy degrees of freedom
and $\hat{H}_S$ denotes the system Hamiltonian. We can write the evolution
operator $\hat{U}$ of the grand system, i.e., the system plus reservoir, as
$\hat{U}=\sum_{E,ij}\ket{\psi_{j,E}} \bra{\psi_{i,E}}s_{ji,E} \hat{F}_{ji,E}$,
where the coefficients $s_{ji,E}$ are elements of the scattering matrix
describing the transitions between the system's quantum states
$\ket{\psi_{i,E}} \rightarrow \ket{\psi_{j,E}}$ (without taking into account
the interaction with the reservoir) and the family of operators
$\hat{F}_{ji,E}$ act in the reservoir Hilbert space, with the subscripts $i$,
$j$, and $E$ specifying the system's states.

To describe the quantum dynamics of an open system, QIT introduces the
so-called \textit{quantum channel} (QC) defined as a trace-preserving
completely positive map, $\hat{\tilde\rho}=\Phi(\hat \rho)$, of a density
matrix $\hat \rho$. To determine whether the evolution belongs to the class
of unital channels, implying that the system evolves with a non-negative
entropy gain $\Delta S=S(\Phi(\hat \rho))-S(\hat \rho)\geq0$, one has to check
if $\Phi(\hat 1)=\hat 1$. Using the unitarity of $\hat U$, one
finds\,\cite{Lesovik}
\begin{equation}
\label{e1}
   \Phi_{jj'}(\hat{1}_E)-[\hat{1}_E]_{jj'}=\sum_i{s_{ji,E} 
   s^*_{j'i,E}\langle[\hat{F}^{\dagger}_{j'i},\hat{F}_{ji}]\rangle},
\end{equation}
where $\langle \ldots \rangle$ denotes averaging with respect to the initial
state of the reservoir and $\hat{1}_E=\sum_i{\ket{\psi_{i,E}}
\bra{\psi_{i,E}}}$. This relation is the central result of Ref.\
\cite{Lesovik}. It allows to formulate the quantum $H$-theorem as follows:
\textit{If the r.h.s. of Equation (\ref{e1}) vanishes, the quantum system
evolves with} $\Delta S\geq0$. Using this result, we have demonstrated that
the typical evolution of energy-isolated quantum systems occurs with
non-diminishing entropy. E.g., we have shown that the electron-phonon
interaction implies that an electron's evolution satisfies the conditions of
the quantum $H$-theorem.  The same is true for an electron interacting with a
random ensemble of three-dimensional (3D) nuclear spins for the case where the
thermodynamic averaging of the commutators in Equation (\ref{e1}) implies
their vanishing.

Yet, we also have uncovered special situations where the Second Law, stating
that the entropy of an isolated system is non-decreasing, can be locally
violated. In \cite{Lebedev}, we have proposed several setups which realize
non-unital and energy-conserving quantum channels and where a
micro-environment acts with two non-commuting operations on the system in an
autonomous way. We have found, that such a process corresponds to a partial
exchange or \textit{swap} between the system's and the environment's quantum
states, with the system's entropy decreasing if the environment's state is
more pure.  This entropy-decreasing process is naturally thought of as the
action of a quantum version of a Maxwell demon. We have proposed a
quantum-thermodynamic engine capable of extracting energy from a single heat
reservoir and perform useful work, provided that pure qubits are available for
the machine's operation.  The special feature of this engine, which involves
an energy-conserving non-unital quantum channel, is the separation of its
entire operation cycle into two subcycles (that are, in principle, spatially
remote), a working cycle and an entropy cycle. This allows the engine to run
with no local waste heat, see Figure \ref{fig1}.

It seems to us, however, that for many reasons it is not really appropriate to
invoke a 'demon' in the operation of this engine, as this expression comes
with many negative connotations and a 'demon's action' may be associated with
certain detrimental work. Our `demon', on the opposite,
serves a useful purpose and, in our view, should instead be called a
\textit{Quantum Angel}. Nevertheless, for the time being, below we use the
traditional terminology.

In the following, we show how our criterion of unitality as expressed in
Equation (\ref{e1}) can be used to identify quantum processes associated with
an entropy change and investigate two examples to which this approach is
relevant. First, we discuss a Maxwell demon based on a qubit, originally
described in Ref.\ \cite{Pekola} within a classical framework. Secondly, we
will consider a process taking place in a two-qubit system that comprises
heating and cooling.

\section{Quantum Version of Maxwell Demon\\ Based on a Single Qubit}
\begin{figure}[h]
  \centerline{\includegraphics[width=420pt]{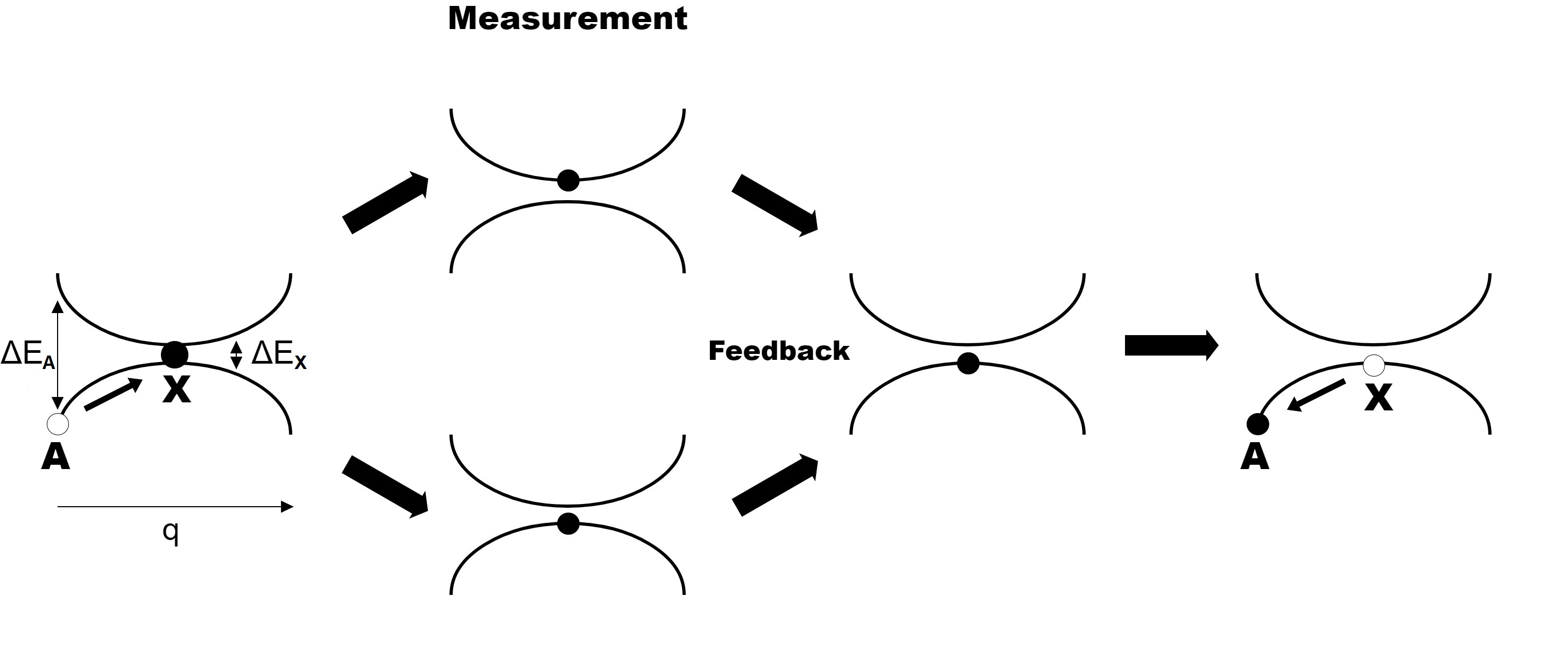}}
  \caption{Schematic operation cycle of a Maxwell demon based on a qubit.}
  \label{fig2}
\end{figure}
In this section, we discuss a system that acts as a Maxwell
demon described in Ref.\,\cite{Pekola}. 
There the engine's operation involves a
standard measurement which is followed by the feedback protocol based on the obtained data. Here, we discuss a fully
quantum version of the demon for which both, the measurement and the conditional feedback can be expressed by
unitary operators. The possibility of such a unitary measuring process was demonstrated in \cite{Oehri}.

The demon's setup is based on a two-level (\textit{qubit}) system that
interacts with the bath characterized by the temperature $T$. The distance
between the energy levels of the qubit ground ($g$) and the exited ($e$)
states can be changed via the control parameter $q$.

At the start of the cycle, the qubit is in the ground state (point A in Fig.
\ref{fig2}) and the level spacing $\Delta E_A \gg k_B T$; thus, the qubit
state is not subject to fluctuations. The control gate is then
changed adiabatically to the value that corresponds to the minimum level
separation $\Delta E_X \ll k_B T$ (point X in Fig.\ \ref{fig2}). This results
in an entropy gain of the qubit by $k_B \ln 2$, with a corresponding amount of
heat $k_B T\ln 2$ being extracted from the bath. Subsequently, another quantum
system (the \textit{manipulator}) performs a measurement of the qubit state.
Based on the obtained information, the manipulator executes a feedback
protocol: if the qubit is in the ground state, $q$ is moved quickly to the
original point A. Otherwise, if the qubit is in the exited state, the
manipulator swaps the qubit state $e\rightarrow g$ by performing the work $-
\Delta E_X$, followed by the transition to the position A.

In our analysis, we focus on the dynamics of the qubit--manipulator compound system only.
In this case, the manipulator plays the role of the Maxwell demon,
so in what follows, we will be using the term 'demon' rather than
'manipulator'.  The time evolution of the compound system during one cycle
can be divided in five stages:
\begin{enumerate}
  \item At time $t_0$, the initial state of the compound system is denoted by
  the point A, its corresponding density matrix is $\hat{R}_i=\hat{\rho}_i
  \otimes \hat{\pi}_i$, where $\hat{\rho}_i$ describes the qubit state, and
  $\hat{\pi}_i=\ket{0} \bra{0}$ describes the demon state.
  \item From $t_0$ to $t_1$, the qubit undergoes a unitary evolution, the
  system transfers from point A to point X, while the demon state remains the
  same. At the time $t_1$, the compound system's state is described by the
  matrix
  \begin{equation}
  \label{e2}
     \hat{R}_1= \hat{\rho}_1\otimes\ket{0} \bra{0},
  \end{equation}
  where the qubit matrix $\hat{\rho}_1$ is diagonal and given by 
  \begin{equation}
  \label{e3}
     \hat{\rho}_1=\rho_{gg} \ket{g} \bra{g}+\rho_{ee} \ket{e} \bra{e},
  \end{equation}
  where $\ket{g}$ stands for the ground state and $\ket{e}$ stands for the
  exited state.
  \item From $t_1$ to $t_2$, the demon receives the information about the
  qubit state. If the qubit is found in the ground
  state, the demon remains in the state $\ket{0}$, otherwise the demon state
  becomes $\ket{1}$ $(\braket{0|1}=0)$. The compound system then evolves
  according to unitary operator
    \begin{equation}
    \label{e4}
        \hat{U}_1=\ket{g} \bra{g} \otimes(\ket{0} \bra{0}+\ket{1} \bra{1})
       +\ket{e} \bra{e} \otimes (\ket{0} \bra{1}+\ket{1} \bra{0}).
    \end{equation}
  At the time $t_2$, the compound system state is
   \begin{equation}
   \label{e5}
      \hat{R}_2=\rho_{gg} \ket{g} \bra{g} \otimes \ket{0} \bra{0}
      +\rho_{ee} \ket{e} \bra{e} \otimes \ket{1} \bra{1}.
  \end{equation}
  \item From $t_2$ to $t_3$, the demon performs a feedback operation on the
  qubit that depends on the obtained information, with a corresponding
  unitary evolution operator 
  \begin{equation}
  \label{e6}
      \hat{U}_2=(\ket{g}\bra{g}+\ket{e}\bra{e})\otimes\ket{0}\bra{0}
      +(\ket{g}\bra{e}+\ket{e}\bra{g})\otimes\ket{1}\bra{1}. 
  \end{equation}
  At the time $t_3$, the compound system's state becomes
  \begin{equation}
  \label{e7}
     \hat{R}_3=\ket{g}\bra{g}\otimes(\rho_{gg} \ket{0} \bra{0}
     +\rho_{ee}\ket{1} \bra{1}).
  \end{equation}
  
   \item From $t_3$ to $t_4$, the qubit ground state reverses to the position
   A through the unitary evolution, while the demon state is being prepared
   for the next cycle.
\end{enumerate}
The vanishing of the terms $\ket{e}\bra{e}\otimes\ket{0}\bra{0}$ and
$\ket{e}\bra{g}\otimes\ket{1}\bra{1}$ of $\hat{U}_2$ after its action on the
compound system's state reflects the fact that the feedback control depends on
the information obtained by the demon. At the same time, these terms
guarantee the unitarity of the evolution.

Let us consider the evolution of the compound system within the framework and
terminology of the quantum $H$-theorem\,\cite{Lesovik}. Namely, we focus on
the third and the fourth stages of the cycle. We note that the compound system
evolution from $t_1$ to $t_3$ is characterized by the unitary operator
\begin{equation}
\label{e8}
  \hat{U} = \hat{U}_2 \hat{U}_1=\ket{g}\bra{g}\otimes\ket{0}\bra{0}
  +\ket{g}\bra{e}\otimes\ket{1}\bra{0}+\ket{e}\bra{g}\otimes\ket{1}\bra{1}
  +\ket{e}\bra{e}\otimes\ket{0}\bra{1}
\end{equation}
(see Equation\,(\ref{e4}) and (\ref{e6})). Ideally, the value of $\Delta E_X$
can be made arbitrary small, which allows us to apply the concept of a
quasi-isolated system to the qubit. The demon, in turn, acts as a reservoir.
For the qubit to decrease its entropy, its evolution must be described by a
non-unital quantum channel ($\Phi$). The general equation (\ref{e1}) assumes
the form
\begin{equation}
\label{e9}
    \Phi_{jj'}(\hat{1})-[\hat{1}]_{jj'}
    =\sum_i{\langle[\hat{F}^{\dagger}_{j'i},\hat{F}_{ji}]\rangle},
\end{equation}
where the operators $\hat{F}_{ji}$ can be obtained from Equation (\ref{e8}):
\begin{align*}
    \hat{F}_{gg}&=\ket{0}\bra{0} & \hat{F}_{ge}&=\ket{1}\bra{0}\\
    \hat{F}_{eg}&=\ket{1}\bra{1} & \hat{F}_{ee}&=\ket{0}\bra{1}.
\end{align*}
Using Equation (\ref{e9}), one finds that 
\begin{equation}
\label{e10}
    \Phi(\hat{1})=2\cdot \ket{g}\bra{g},
\end{equation}
which proves the non-unitality of the quantum channel.

\section{Heating and Cooling Process}

In this section, we address heating and cooling processes that may occur in
complex physical systems. Instead of
describing fully realistic processes we want to represent them on a formal level using unitary evolution of two interacting
qubits. Below we will consider both heating and cooling operations
within this perspective.

Suppose that the Hilbert space of each qubit has an orthonormal basis which
consists of $\ket 0$ and $\ket 1$. Let the initial state of the two-qubit system 
be described by the density matrix
\begin{equation}
\label{e11}
    \hat{R}_i=\ket{0}\bra{0}\otimes\frac{1}{2}(\ket{0}\bra{0}+\ket{1}\bra{1}).
\end{equation}
The matrix of the final state after the first qubit's heating and the second
qubit's cooling is given by
\begin{equation}
\label{e12}
    \hat{R}_f=\frac{1}{2}(\ket{0}\bra{0}+\ket{1}\bra{1})\otimes\ket{0}\bra{0},
\end{equation}
i.e., the first qubit changes its state to chaotic while the second qubit goes
to the ground state. The system evolution in such a process can be expressed
via the unitary operator
\begin{equation}
\label{e13}
  \hat{U} =\ket{0}\bra{0}\otimes\ket{0}\bra{0}+\ket{0}\bra{1}
  \otimes\ket{1}\bra{1}+\ket{1}\bra{0}\otimes\ket{0}\bra{1}
  +\ket{1}\bra{1}\otimes\ket{1}\bra{0}.
\end{equation}
Let us now analyze the evolution of each qubit, i.e., decompose the process into the first qubit heating and the
second qubit cooling.

Since the heating process is associated with an increase in entropy, the
quantum channel $\Phi^{(1)}$ describing the evolution of the first qubit may
be unital.
The general formula (\ref{e1}) takes the form $\Phi^{(1)}_{jj'}(\hat{1}) -
[\hat{1}]_{jj'} = \sum_i{\langle[\hat{F}^{(1)\dagger}_{j'i},
\hat{F}^{(1)}_{ji}]\rangle}$, where the operators $\hat{F}^{(1)}_{ji}$ can be
obtained from Equation\,(\ref{e13}):
\begin{align*}
    \hat{F}^{(1)}_{00}&=\ket{0}\bra{0} & \hat{F}^{(1)}_{01}&=\ket{1}\bra{1}\\
    \hat{F}^{(1)}_{10}&=\ket{0}\bra{1} & \hat{F}^{(1)}_{11}&=\ket{1}\bra{0}.
\end{align*}
Consequently, we find that 
\begin{equation}
\label{e14}
    \Phi^{(1)}(\hat1)=\hat1,
\end{equation}
implying that $\Phi^{(1)}$ is indeed unital as could be expected.

The cooling process involves a decrease of entropy, which means that the
quantum channel $\Phi^{(2)}$ corresponding to the second qubit has to be
non-unital.  As in the previous case,
$\Phi^{(2)}_{jj'}(\hat{1})-[\hat{1}]_{jj'} = \sum_i
{\langle[\hat{F}^{(2)\dagger}_{j'i}, \hat{F}^{(2)}_{ji}]\rangle}$, however,
now
\begin{align*}
    \hat{F}^{(2)}_{00}&=\ket{0}\bra{0} & \hat{F}^{(2)}_{01}&=\ket{1}\bra{0}\\
    \hat{F}^{(2)}_{10}&=\ket{1}\bra{1} & \hat{F}^{(2)}_{11}&=\ket{0}\bra{1}.
\end{align*}
On substitution, we find that 
\begin{equation}
\label{e15}
    \Phi^{(2)}(\hat{1})=2\cdot\ket{0}\bra{0},
\end{equation}
which proves that $\Phi^{(2)}$ is non-unital as expected.

\section{Concluding Remarks}
In summary, using the results of quantum information theory, we have extended
the Second Law of Thermodynamics to the quantum world. We have derived a
rigorous general criterion of unitality and have applied it to develop the
concept of a quantum Maxwell demon and to describe the heating--cooling process
in a two-qubit system.

\section*{Acknowledgments}

The research was supported by the Government of the Russian Federation
(Agreement № 05.Y09.21.0018), by the RFBR Grants No. 17- 02-00396A and
18-02-00642A, Foundation for the Advancement of Theoretical Physics "BASIS",
the Ministry of Education and Science of the Russian Federation
16.7162.2017/8.9 (A.V.L., in part related to the section 2), the Swiss
National Foundation via the National Centre of Competence in Research in
Quantum Science and Technology (NCCR QSIT), and the Pauli Center for
Theoretical Physics.  The work of V.M.V. was supported by the U.S. Department
of Energy, Office of Science, Materials Sciences and Engineering Division.

\nocite{*}
\bibliographystyle{aipnum-cp}%
\bibliography{sample}%

\end{document}